\documentclass[a4paper]{jpconf}
\usepackage{graphicx}
\begin{document}
\title{Nuclear physics uncertainties of the astrophysical $\gamma$-process studied through the $^{64}$Zn(p,$\alpha$)$^{61}$Cu and $^{64}$Zn(p,$\gamma$)$^{65}$Ga reactions}

\author{Gy. Gy\"urky, Zs. F\"ul\"op, Z. Hal\'asz, G. G. Kiss, T. Sz\"ucs}
\address{Institute for Nuclear Research (Atomki), Bem t\'er 18/c, H-4026 Debrecen, Hungary}

\ead{gyurky@atomki.mta.hu}

\begin{abstract}

In a recent work, the cross section measurement of the $^{64}$Zn(p,$\alpha$)$^{61}$Cu reaction was used to prove that the standard $\alpha$-nucleus optical potentials used in astrophysical network calculation fail to reproduce the experimental data at energies relevant for heavy element nucleosynthesis. In the present paper the analysis of the obtained experimental data are continued by comparing the results with the predictions using different parameters. It is shown that the recently suggested modification of the standard optical potential leads to a better description of the data. 

\end{abstract}

\section{Introduction}

Nature builds up the chemical elements of the universe through nuclear reactions starting from the nucleons that were born after the Big Bang. We need to understand these nuclear reactions in order to explain the abundances and the mere existence of chemical elements. The mechanism of nuclear reactions is therefore a key ingredient in our understanding of nucleosynthesis. Ab initio calculations based on the basic nucleus-nucleus interaction are possible only for the lightest systems of a few nucleons. For the reactions of heavier nuclei simplifications are necessary such as the reduction of the many body problem of a reaction to a two body one. In such a picture the interaction of the two reacting nuclei are represented by a nucleus-nucleus optical potential. The properties of such a potential - among other factors - determine the calculated cross section. It is therefore essential to find a suitable potential in order to obtain reliable cross sections for astrophysical reaction network calculations of the elemental and isotopic abundances.

The appropriate potential acting between interacting nuclei plays an important role in practically all mass regions encountered in nuclear astrophysics from low level density systems (where e.g. direct capture cross sections are calculated by potential models) to high level densities towards the heavier mass region. The motivation of the present work came from the problems related to the synthesis of the heavy proton rich isotopes (p-isotopes) in the astrophysical $\gamma$-process \cite{rau13}. In the mass and energy region of the $\gamma$-process nuclear reactions proceed through the formation of compound nuclei at excitation energies where the level density is high. Nuclear reactions in this domain are usually described using the statistical approach of the Hauser-Feshbach theory. An important ingredient of the calculations is the averaged transmission coefficients in all energetically allowed reaction channels. In the case of particle emission or absorption, the transmission coefficients are obtained from the corresponding particle-nucleus optical potential (where the particle can be a single nucleon or a few nucleon system like an $\alpha$-particle).

Owing to the generally unsatisfactory results of $\gamma$-process network calculations, the experimental study of the relevant nuclear reactions became a hot topic of nuclear astrophysics research of recent years \cite{gyu15}. Many experimental results indicate that statistical model calculations are not able to describe well the measured cross sections of reactions involving $\alpha$-particles. One of the most likely reasons of the failure is the lack of good knowledge of the $\alpha$-nucleus optical potential at low energies. This was identified typically in measurements of $\alpha$-induced reactions \cite{szu14}. The problem with these reactions, however, is the fact that owing to the extremely low cross section the measurements are done at energies higher than the astrophysically relevant ones. This means that the $\alpha$-nucleus optical potential is not tested in the relevant energy range. Moreover, the cross sections at the studied energies show sensitivity not only to the $\alpha$-nucleus potential, but to several other parameters. In order to study directly the $\alpha$-nucleus potential at astrophysical energies, a new approach is followed in the present work.

\section{The studied reactions and the experimental technique}

In order to study the low energy $\alpha$-nucleus optical potential at $\gamma$-process relevant energies, the cross section of a reaction where the $\alpha$-particle is in the exit channel was measured. The chosen reaction was $^{64}$Zn(p,$\alpha$)$^{61}$Cu. The advantages of this reaction and the experimental technique was detailed in our recent publication \cite{gyu14}. Here only the most important aspects are summarized.

Calculations show that in the proton energy range between about 3 and 8\,MeV the cross section of this reaction obtained in the statistical model depends solely on the choice of the $\alpha$-nucleus optical potential \cite{rau12}. This is indicated as a sensitivity plot shown in figure\,1 of ref \cite{gyu14}. Taking into account the Q value of the reaction, the $\gamma$-process relevant energy range of the $^{61}$Cu\,+\,$\alpha$ system in the exit channel of the studied reaction lies in the lower half of this energy interval. Therefore the measurement of the $^{64}$Zn(p,$\alpha$)$^{61}$Cu cross section provides direct information about the $\alpha$-nucleus potential.

The experiments were carried out at the cyclotron accelerator of Atomki where $^{64}$Zn targets were bombarded by a proton beam in the energy interval between 3.5 and 8\,MeV. Since the reaction product is radioactive, the number of reactions was determined by the activation method measuring the $\gamma$-radiation following the decay of $^{61}$Cu by a HPGe detector.

The most important conclusion of the measurements was already drawn in ref \cite{gyu14}. Namely, the experiments proved for the first time at astrophysical energies that the commonly used optical potential strongly overestimates the measured cross sections towards the lowest energies. This overestimation is a general finding of earlier experiments, which were however carried out at much higher $\alpha$\,+\,nucleus relative energies. In this paper some further comparison of the measured cross sections with model calculations is shown. For the comparisons two statistical model codes are used. The SMARAGD code \cite{SMARAGD} is a successor of the NON-SMOKER code \cite{NONSMOKER} which is often applied in nuclear astrophysics network calculations. Calculations with the widely used TALYS \cite{TALYS} code are also carried out.


\section{The $^{64}$Zn(p,$\gamma$)$^{65}$Ga reaction, $\gamma$-strength functions and level densities}

\begin{figure}[t]
\centering
\includegraphics[angle=-90,width=0.49\textwidth]{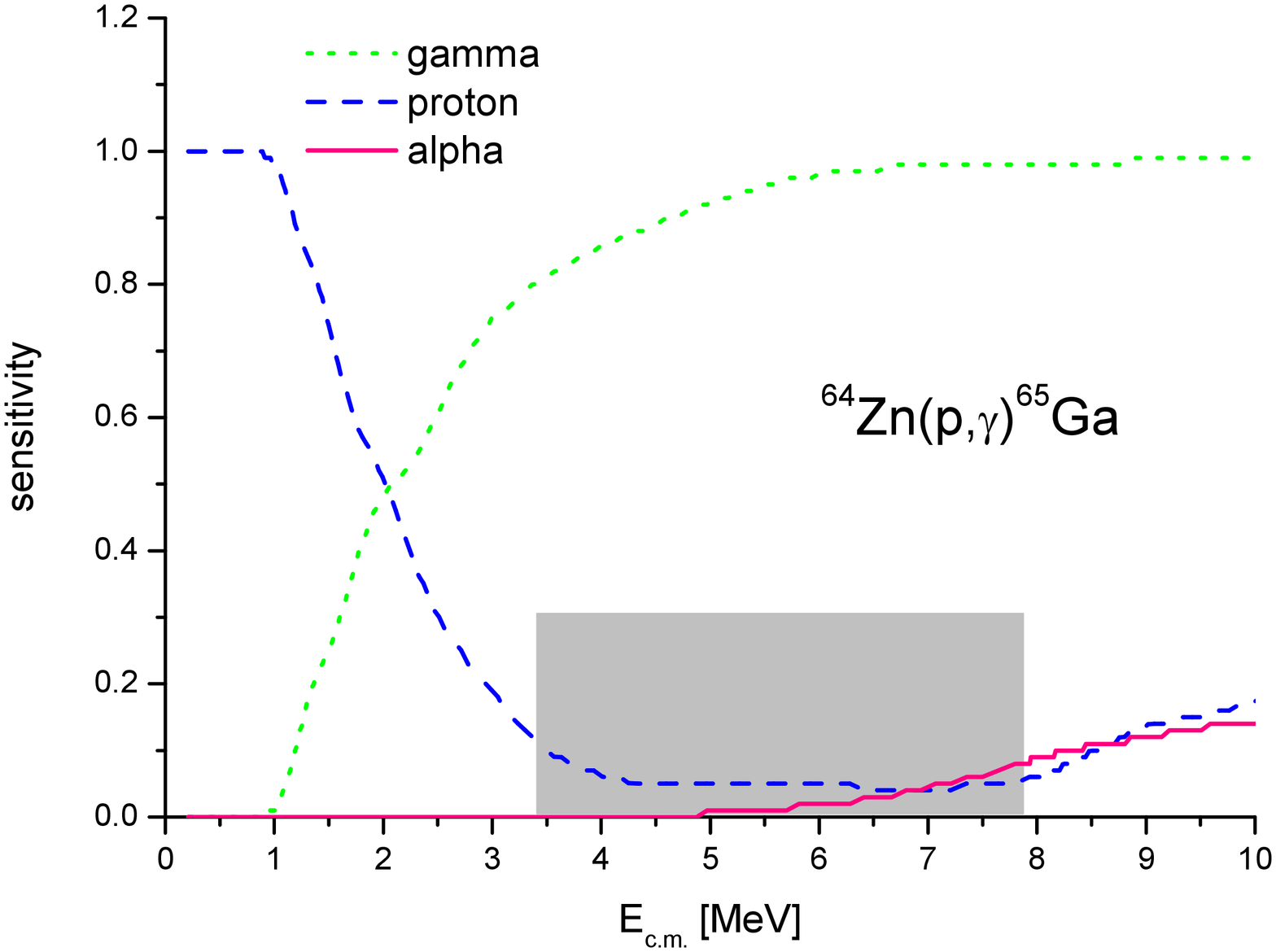}
\includegraphics[angle=-90,width=0.49\textwidth]{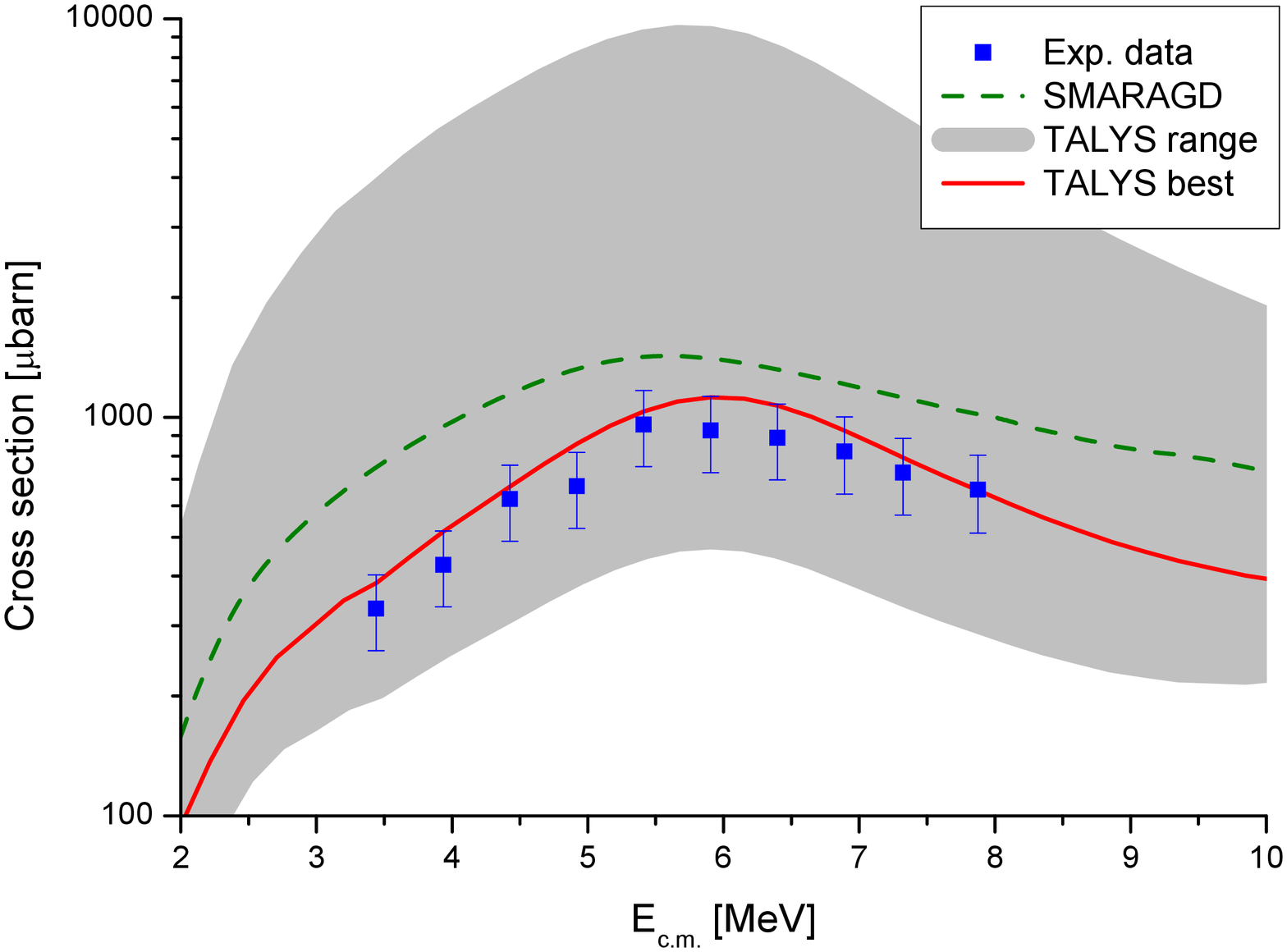}
\caption{\label{fig:pg} Left: sensitivity of the calculated $^{64}$Zn(p,$\gamma$)$^{65}$Ga cross section to the various partial widths. Right: measured and calculated $^{64}$Zn(p,$\gamma$)$^{65}$Ga cross section. See text for details.}
\end{figure}

Although the primary aim of the present work was the measurement of the $^{64}$Zn(p,$\alpha$)$^{61}$Cu cross section, the $^{64}$Zn(p,$\gamma$)$^{65}$Ga reaction was also studied as this is the other open reaction channel on $^{64}$Zn in the studied energy range. Similarly to figure\,1 of ref \cite{gyu14}, the left panel of figure\,\ref{fig:pg} here shows the sensitivity of the  $^{64}$Zn(p,$\gamma$)$^{65}$Ga calculated cross section to the variation of different partial widths. As one can see, the (p,$\gamma$) cross section in the studied energy range (shaded area) is sensitive solely to the $\gamma$-width. 

The right panel of figure\,\ref{fig:pg} shows the measured $^{64}$Zn(p,$\gamma$)$^{65}$Ga cross section and the predictions of the statistical model calculations. The standard SMARAGD calculation overestimates the measured data by about a factor of two but describes relatively well the energy dependence. The $\gamma$-width is determined by the level density and $\gamma$-ray strength function used in the statistical model. TALYS calculations were therefore carried out with varying these input parameters. For both parameters TALYS version 1.4 used in this work offers five different built-in options. Calculations with all possible combinations of these parameters were done and the cross section range covered by the obtained results is shown in the figure as the gray band. As one can see, differences of up to a factor a 50 can be obtained with the different input parameters. It is worth noting that there is no a priori information about which of the possible input parameters are correct. The wide range of possible calculated cross sections show that the not correct nuclear reaction rates entering the $\gamma$-process models can indeed be the reason of the poor reproduction of the p-isotope abundances. Only experiments like the present one can help find the best input parameters for the calculations. In the case of the $^{64}$Zn(p,$\gamma$)$^{65}$Ga reaction the best TALYS result can be obtained using the microscopic level density of Goriely \textit{et al.} \cite{gor01} and the $\gamma$-strength function of Kopecky and Uhl (generalized Lorentzian form of the giant dipole resonance for the dominant E1 transition) \cite{kop90}. This is shown as the red curve in figure\,\ref{fig:pg}.

\section{The $^{64}$Zn(p,$\alpha$)$^{61}$Cu reaction and the modification of the $\alpha$-nucleus potential}

\begin{figure}[t]
\centering
\includegraphics[angle=-90,width=0.53\textwidth]{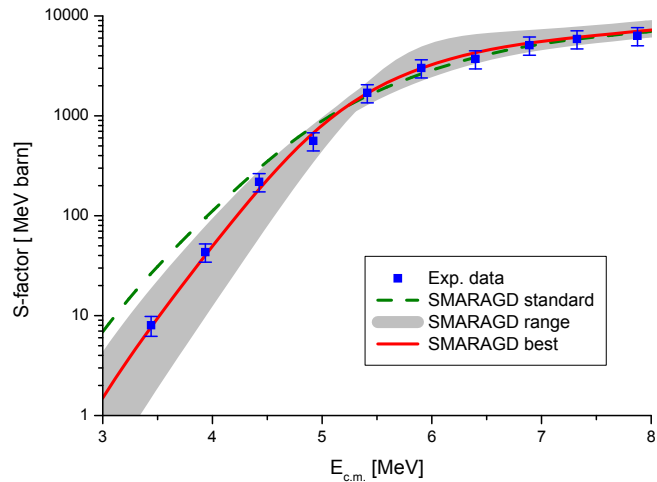}
\caption{\label{fig:pa} Measured and calculated astrophysical S-factor of the $^{64}$Zn(p,$\alpha$)$^{61}$Cu cross section. See text for details.}
\end{figure}

The most important conclusion of our measurement presented in ref\,\cite{gyu14} was that the standard statistical model calculations overestimate the measured $^{64}$Zn(p,$\alpha$)$^{61}$Cu at the lowest studied energies while the agreement becomes fairly good when the energy increases. In the case of the SMARAGD code (and in its predecessor NON-SMOKER) the standard calculations are carried out with the alpha-nucleus optical potential of McFadden and Satchler \cite{mcf66}. This potential was developed based on 25\,MeV experimental data and does not have any energy dependence. It is therefore not surprising that experimental evidence is gathering that the potential fails at low energies where an energy dependence of at least the imaginary part of the potential seems to be required. 

In order to cure the low energy problem of the McFadden--Satchler potential, an energy dependent modification was recently recommended \cite{sau11}. The depth of the imaginary part was modified using a Fermi-type function with a diffuseness parameter $a_E$. Several recent experiments indicate that this modification leads to a better reproduction of the measured data (see e.g. \cite{yal15} and references therein). In the present paper this modification is tested for the $^{64}$Zn(p,$\alpha$)$^{65}$Cu reaction. Figure \ref{fig:pa} shows the comparison of the experimental data with SMARAGD calculation in the form of astrophysical S-factor. (Calculations with the TALYS code are omitted here as they were presented in \cite{gyu14}.) The green line shows the standard calculation using the original McFadden--Satchler potential. The grey band indicates the obtained S-factors by the modified potential with the $a_E$ parameter in the range between 1.5 and 5 MeV. It is evident that the modification leads to a much better reproduction of the experimental data. The best agreement is achieved with $a_E=2.5$\,MeV as indicated by the red curve.

\section{Summary and conclusions}

The present experimental data prove directly at astrophysical energies that the standard low energy $\alpha$-nucleus optical potential must be modified. The new data support the recently suggested modification of the potential. The implementation of this modification in the reaction rate calculations for astrophysical models is recommended. The comparison of the measured and calculated $^{64}$Zn(p,$\gamma$)$^{65}$Ga cross section data indicate that further experimental effort is needed as only measured data can help find the correct nuclear input parameters of calculations which show highly divergent results depending on the parameters.

\ack The authors thank T. Rauscher for providing the results of the SMARAGD code. This work was supported by OTKA grants No. K101328 and K108459.

\end{document}